# High-fidelity, high-isotropic resolution diffusion imaging through gSlider acquisition with $B_1^+$ & $T_1$ corrections and integrated $\Delta B_0$/Rx shim array

Congyu Liao[1,2*], Jason Stockmann[1,2], Qiyuan Tian[1,2], Berkin Bilgic[1,2], Nicolas S. Arango[1,3], Mary Kate Manhard[1,2], William A. Grissom[4], Lawrence L. Wald[1,2], and Kawin Setsompop[1,2]

[1] Athinoula A. Martinos Center for Biomedical Imaging, Massachusetts General Hospital, Charlestown, MA, USA.

[2] Department of Radiology, Harvard Medical School, Boston, MA, USA.

[3] Department of Electrical Engineering and Computer Science, Massachusetts Institute of Technology, Cambridge, MA, USA

[4] Institute of Imaging Science, Vanderbilt University, Nashville, TN, USA.

* <u>Corresponding author:</u>
Congyu Liao, PhD, CLIAO2@mgh.harvard.edu, Athinoula A. Martinos Center for Biomedical Imaging, Building 149, Room 2301, 13th Street, Charlestown, MA, 02129 USA

# Abstract

**Purpose:**

$B_1^+$ and $T_1$ corrections and dynamic multi-coil shimming approaches were proposed to improve the fidelity of high isotropic resolution Generalized slice dithered enhanced resolution (gSlider) diffusion imaging.

**Methods:**

An extended reconstruction incorporating $B_1^+$ inhomogeneity and $T_1$ recovery information was developed to mitigate slab-boundary artifacts in short-TR gSlider acquisitions. Slab-by-slab dynamic $B_0$ shimming using a multi-coil integrated $\Delta B_0$/Rx shim-array, and high in-plane acceleration ($R_{inplane}=4$) achieved with virtual-coil GRAPPA were also incorporated into a 1 mm isotropic resolution gSlider acquisition/reconstruction framework to achieve an 8-11 fold reduction in geometric distortion compared to single-shot EPI.

**Results:**

The slab-boundary artifacts were alleviated by the proposed $B_1^+$ and $T_1$ corrections compared to the standard gSlider reconstruction pipeline for short-TR acquisitions. Dynamic shimming provided >50% reduction in geometric distortion compared to conventional global 2nd order shimming. 1 mm isotropic resolution diffusion data show that the typically problematic temporal and frontal lobes of the brain can be imaged with high geometric fidelity using dynamic shimming.

**Conclusions:**

The proposed $B_1^+$ and $T_1$ corrections and local-field control substantially improved the fidelity of high isotropic resolution diffusion imaging, with reduced slab-boundary artifacts and geometric distortion compared to conventional gSlider acquisition and reconstruction. This enabled high-fidelity whole-brain 1 mm isotropic diffusion imaging with 64 diffusion-directions in 20 minutes using a 3T clinical scanner.

# Introduction

High-resolution diffusion-weighted imaging (DWI) with echo-planar imaging (EPI) acquisition technique is a powerful tool for many neuroscientific and clinical applications. Single-shot EPI (SS-EPI) (1) is one of the most commonly used methods in DWI due to its rapid encoding ability. However, the SS-EPI readout does not lend itself to high in-plane resolution imaging due to severe geometric distortion and $T_2^*$-related blurring artifacts, which are difficult to correct in post-processing, particularly where voxel pile-up occurs (2). Multi-shot EPI (3–5) is a promising approach to improve the geometric fidelity of DWI and achieve high in-plane resolution with low distortion and $T_2^*$ blurring. However, due to the prolonged scan time, shot-to-shot phase variations and potential patient motion, multi-shot EPI continues to be a challenge in DWI. To mitigate these problems, previous studies combined multi-shot approaches with parallel imaging (6,7), sparse or low-rank models (8,9), joint reconstruction (10–12) and simultaneous multi-slice (SMS) (13–15) to accelerate the acquisition and correct for shot-to-shot phase variations.

Another approach to mitigate distortion is to generate a compensating $B_0$ magnetic field to counteract the off-resonance effects. Conventional scanners are equipped with $1^{st}$ order (the linear gradients) and static $2^{nd}$ order spherical harmonic shim coils, which generate a spatial magnetic profile to compensate the $B_0$ field over the target volume (16,17). Local multi-coil (MC) shimming using small shim coils patterned around the imaged object has been introduced as a convenient way to provide higher-order $B_0$ shimming without the need to substantially modify the MRI scanner (18–21). Compared to $2^{nd}$ order shimming, MC shim arrays have been shown to provide improved $B_0$ homogeneity to improve EPI acquisitions for structural and fMRI applications (22,23), and magnetic resonance spectroscopic imaging (MRSI) acquisition for spectroscopy studies (24). The ability of MC arrays to rapidly switch their shim currents without causing image artifacts allows the $B_0$ shim to be optimized on a slice-by-slice basis, providing further gains in $B_0$ homogeneity and mitigation of EPI geometric distortion (18).

Improving SNR efficiency is critical for achieving high isotropic resolution DWI. Three dimensional multi-slab DWI has emerged as a promising strategy to enhance the SNR in such acquisitions (25–29). However, shot-to-shot phase variations and slab-boundary artifacts are still key challenges for efficient sampling of whole-brain high-resolution DWI with this technique (30). Generalized SLIce Dithered Enhanced Resolution (gSlider) is a simultaneous multi-slab (SMSb) acquisition technique with self-navigated RF slab-encoding, which has shown great potential for motion-robust, high-resolution DWI (31,32). Here, the simultaneous multi-slabs are acquired together using the blipped-CAIPI acquisition scheme (33) and separated through parallel imaging. The encoding within each slab is then performed through sequential RF slab-encoding acquisitions and combined using super-resolution reconstruction to create high-resolution slices. The gSlider RF-encodings are designed to provide high signal in each of the RF-encoded acquisitions for robust estimation and removal of shot-to-shot phase corruptions without the need for an additional navigator. With gSlider, a large number of slices can be acquired together per EPI-encoding (e.g. 10 simultaneous slices using gSlider RF-encoding of 5x and Multiband factor of 2x) to achieve a short-TR for high resolution volumetric DWI and provide high SNR-efficiency. With this approach, the ratio of slab thickness relative to slice resolution can be kept small (e.g. 5 for gSlider 5x) which allows for sharp slab selective excitation with reduced slab boundary issues compare to typical multi-slab DWI. This has enabled high-quality submillimeter gSlider DWI without slab boundary correction at TR ~4.5s or above (31,32,34). However, at shorter TRs slab boundary artifacts can still remain an issue, with partial $T_1$ recovery in adjacent slabs causing striping artifacts.

In this work, we developed approaches to improve the fidelity of gSlider: first by making it robust to slab-boundary artifacts, and second by reducing geometric distortion from $B_0$ inhomogeneity. To mitigate slab-boundary artifacts in short TR acquisitions, dictionary-based $B_1^+$ inhomogeneity and $T_1$ recovery corrections were incorporated into gSlider reconstruction. To mitigate image distortions, a 32-channel integrated $\Delta B_0$/Rx array (AC/DC coil) (18) was utilized for slab-by-slab shimming, to reduce $B_0$ inhomogeneity by >50% as compared to static $2^{nd}$ order shimming. This dual-purpose coil array provides both high spatial-order $B_0$ field control as well as good parallel imaging capability. This is then combined with high in-plane acceleration ($R_{inplane}$=4) and virtual conjugate coil (VCC-) GRAPPA reconstruction (10,35), to achieve 8-11 fold total geometric distortion reduction in single-shot gSlider-EPI. We demonstrate that the proposed method can achieve high-fidelity whole-brain 1 mm isotropic DWI with 64 diffusion-directions in ~20 minutes on a clinical 3T scanner.

## Methods

**Pulse sequence design and slab-optimized dynamic shimming**

Figure 1(a) shows the sequence diagram of gSlider, where two external triggers are added in each TR to enable slab-by-slab $B_0$ shimming with the AC/DC coil (Fig 1(d)). To avoid poor performance in whole-brain fat suppression from large out-of-slab $B_0$ inhomogeneity, the slab-by-slab shimming was turned off during fat saturation. In each TR, an additional $k_y$ blip was added to shift k-space and create more unique source points for improved virtual conjugate-coil GRAPPA reconstruction at high accelerations in DWI as outlined in (10,36).

To generate a sharp slab-encoding performance, the Shinnar-Le Roux (SLR) algorithm (37) was used to design the gSlider excitation RF pulses. The time-bandwidth product of the 90º gSlider RF excitation pulses was 12, with a pulse duration of 11 ms. Figure 1(b) and (c) show the waveforms of five gSlider-encoding RF pulses that were used and their corresponding Bloch simulated slab profiles. After slab-encoding, a standard SLR spin-echo (SE) refocusing pulse was applied *without* gSlider-encoding. The time-bandwidth product of refocusing pulse was 8 and the duration of the refocusing pulse was 7.3 ms. The blue lines in Fig.1(c) show the sub-slab profiles of gSlider-encoding after the SE refocusing pulse. To reduce the peak power of the RF pulses, a VERSE method (38) was applied to both gSlider excitation RF pulses and SE refocusing pulse.

To implement slab-optimized shimming, a low-resolution $B_0$ field-map with conventional global-shims applied was acquired using a vendor-provided two-TE gradient echo field mapping sequence. The field maps were registered to the thin-slab gSlider images and then masked using FSL BET and phase unwrapped using FSL PRELUDE (39). The optimal DC shim currents in each channel of the AC/DC coil were then computed on a slab-wise basis using a previously-acquired calibration $B_0$ map basis set for the array. The details of the constrained optimization algorithm can be found in (18). For SMSb acquisition, the shims were jointly optimized over the two simultaneously-acquired slabs. To match the geometric distortions between slab-collapsed EPI data and fully-sampled reference data, the slab ordering of shimming and the currents used in the individual slab of reference data are the same as the corresponding slab group of the SMSb data. To avoid artifacts, the GRAPPA and SMS calibration scan data for each slab were shimmed with the same MC shim fields as the DWI acquisition.

**$B_1^+$ and $T_1$ corrections for robust gSlider reconstruction**

To eliminate shot-to-shot background phase variations in the acquired diffusion data, real-valued diffusion processing (40) was applied. gSlider reconstruction was then performed to obtain high slice-resolution data, using a forward model based on the Bloch simulated slab profiles of the gSlider encodings. Pseudoinverse with Tikhonov regularization was used:

$$\mathbf{X}=(\mathbf{A}^T\mathbf{A}+\lambda\mathbf{I})^{-1}\mathbf{A}^T\mathbf{b} , \qquad [1]$$

where $\mathbf{b}$ (matrix size: $(N_{slab} \cdot N_{rf\text{-}encoding}) \times 1$) is the concatenation of acquired thin-slab data at a given in-plane spatial location, $\mathbf{X}$ (matrix size: $N_{slice} \times 1$) is the corresponding super-resolution reconstruction, $\mathbf{A}$ (matrix size: $(N_{slab} \cdot N_{rf\text{-}encoding}) \times (N_{slab} \cdot N_{rf\text{-}encoding})$) is the RF-encoding matrix that contains the sub-slab profiles simulated from the Bloch equations and $\lambda$ is a Tikhonov regularization parameter. In our previous work (31), the same slab profiles ($M_{xy}$) of the RF-encodings were used for the reconstruction at all spatial locations contained within the A encoding-matrix. However, this does not account for potential spatial variations in $M_{xy}$ due to $B_1^+$ inhomogeneity. Furthermore, the initial longitude magnetization ($M_z$) in the Bloch simulation was set to 1, which ignored incomplete $T_1$ recovery. This can be particularly problematic in the adjacently partial excitation regions of a non-ideal RF excitation after TR/2 slab-interleaved acquisition at short TRs. These imperfections could cause slab-boundary artifacts due to $B_1^+$ variations.

To mitigate slab-boundary artifacts, RF-encoding imperfections due to $B_1^+$ inhomogeneity and incomplete $T_1$ recovery were estimated and incorporated into the RF-encoding matrix **A** of the gSlider reconstruction in Eq.[1]. Figure 2(a) shows the flowchart of $B_1^+$ inhomogeneity correction, where RF-encoding profiles at a range of discretized $B_1^+$ values ([0.70:0.05:1.30], ±30% $B_1^+$ variations) are simulated by using Bloch equation, which enabled the creation of a dictionary of RF-encoding matrices with different $B_1^+$ variations. The corresponding RF-encoding matrix from the dictionary was then selected in each spatial location based on a discretized $B_1^+$ map, which is shown in Fig.2(c).

For incomplete $T_1$ recovery, non-ideal slab-profiles of RF-encodings can cause partial excitations in adjacent slabs which are not fully-recovered in slab-interleaved acquisitions with short TRs. This effect was also modeled by adding partial recovered initial longitude magnetizations $M_z$ into the Bloch simulation of the RF-encodings (assuming average $T_1$=1000 ms in the brain), thereby incorporating them into the encoding matrix. Figure 2(b) shows the partial $M_z$-recovery from adjacent slabs excitations before and after TR/2 longitudinal-relaxation at various $B_1^+$ excitation levels.

**Data acquisition**

All *in vivo* measurements were performed on a 3T scanner (MAGNETOM Prisma, Siemens Healthineers, Erlangen, Germany) with a custom 32-channel AC/DC receive array with added $B_0$ shim capability (18). To assess the improvements provided by slab-optimized shimming, gSlider-EPI with 5 slab-encodings and the corresponding $B_0$ field maps were acquired. The imaging parameters for gSlider data were: FOV 220×220×170

mm$^3$, 34 thin-slabs (5 mm slab-encoding), TR/TE=5100/77 ms, and echo spacing=0.93 ms. To accentuate changes in geometric distortion, data were acquired using both Anterior-to-Posterior (AP) and Posterior-to-Anterior (PA) phase-encodings at different in-plane accelerations ($R_{inplane}$=1 and $R_{inplane}$ =4), with and without slab-optimized shimming. The $B_0$ field maps were acquired using two-echo GRE with 2.5mm slice-thickness and 100% gap. The slice resolution including gap matches the 5 mm gSlider slab-encoding. To validate the shim performance of SMSb imaging, the same gSlider data were acquired with a multi-band (MB) factor of 2 and compared with non-SMSb gSlider data. A matching $T_2$ turbo spin-echo ($T_2$-TSE) data was also acquired as a distortion-free reference.

Whole-brain 1 mm isotropic resolution diffusion imaging data were also acquired with gSlider-EPI and dynamic MC shimming. The protocol used: FOV 220×220×170 mm$^3$, $R_{inplane}$× MB×gSlider =4×2×5, 34 thin-slabs (5 mm slab-encoding), b = 1000 s/mm$^2$ with 64 diffusion-directions and 4 interleaved b =0 s/mm$^2$, TR/TE =3500/86 ms. The total acquisition time is ~20 minutes.

To correct the $B_1^+$ effects in gSlider data, an FOV-matched $B_1^+$ map was obtained by using a Turbo-FLASH scan with pre-conditioning RF pulses (41). The in-plane resolution is 3.4×3.4 mm$^2$ with 2.5 mm slice thickness and 100% gap. The slice resolution including gap matched the 5 mm gSlider slab-encoding.

**Reconstruction and post-processing**

To enable higher in-plane acceleration compared to conventional slice- and in-plane-GRAPPA reconstruction and further reduce the geometric distortions, virtual conjugate-coil GRAPPA with phase-matching was used to achieve high fidelity reconstruction for high acceleration factors. To achieve faster GRAPPA reconstruction, SVD coil compression (42) was applied to compress the 32 channel coil to 20 channels. After GRAPPA reconstruction, the five RF-encoded volumes of each diffusion-direction were then combined to create thin-slice data, using gSlider reconstruction with and without the proposed modified RF-encoding matrix. The RF-encoding matrix was generated using the SLR RF-pulse design and Bloch simulation toolbox (43) (https://vuiis.vumc.org/~grissowa/software.html). The virtual conjugate-coil GRAPPA and gSlider reconstruction algorithms were implemented in MATLAB R2014a (The MathWorks, Inc., Natick, MA). The reconstructed data were then corrected for motion and eddy-current distortion using the "eddy_correct" function from the FMRIB Software Library (39) (FSL, https://fsl.fmrib.ox.ac.uk/fsl/fslwiki/). Diffusion tensor model was fitted using FSL's "dtifit" function to obtain the fractional anisotropy (FA) maps and the primary eigenvectors.

## Results

Figure 3 shows the results of $B_1^+$ and $T_1$ corrections in gSlider reconstruction at TR of 3.5s. With the $B_1^+$ correction only, the slab-boundary artifacts shown in the sagittal and coronal views of a diffusion-weighted volume are well mitigated compared to the standard gSlider reconstruction without corrections. As the white arrow highlights in Fig.3, incorporating $T_1$ correction into the $B_1^+$ corrected processing can further reduce the slab-boundary artifacts, which demonstrates the utility of the proposed $B_1^+$ and $T_1$ corrections in gSlider reconstruction in a short-TR acquisition.

Figure 4 shows compares image distortion with and without dynamic slab-optimized MC $B_0$ shimming. The green arrows highlight the $B_0$ distortion that was alleviated, with slab-by-slab shimming achieving >50% reduction of standard deviation (std) in $\Delta B_0$ across the slab when compared to baseline global 2$^{nd}$ order shimming. Dynamic MC shimming was then combined with in-plane acceleration ($R_{inplane}$=4) to achieve an 8-11x total reduction in $\Delta B_0$ distortion (depending on the slab), yielding images with outlines (red outlines in Fig.4) closely matching that of the reference $T_2$-TSE images.

Figure 5 shows the $B_0$ field maps obtained from global 2$^{nd}$ order shim, MB-2 and MB-1 dynamic MC shimming. Compared to MB-1 slab-optimized shimming, the MB-2 case applies shims simultaneously to two *distant* slabs, providing similar standard deviations of $\Delta B_0$ variations to the MB-1 case, while the out-of-slab regions were unconstrained and allowed to have a poor shim.

Figure 6 shows the dynamic MC shimming results of two representative slabs for MB-1 and MB-2 with A-P and P-A phase encoding directions. Compared to the reference images, the contours of both the MB-1 and MB-2 images closely match those of the $T_2$-TSE images, which demonstrates that the dynamic MC shimming of MB-2 achieved a similar performance to MB-1 slab-optimized dynamic MC shimming. This demonstrates that the 32ch AC/DC coil has enough degrees of freedom to simultaneously control the $B_0$ field in two spatially-separated slices at the same time with minimal loss of performance.

Figure 7 shows the averaged DW images from 64 diffusion-encoding directions (Fig.7a) and directionally-encoded color FA maps (Fig.7b) of the 1mm isotropic gSlider diffusion data. High-quality results are shown in Fig.7(c), with minimal geometric distortions in the typically problematic temporal and frontal lobes. The high-quality results depict the primary eigenvectors from DTI with high fidelity (Fig. 7c) which are beneficial for mapping structural connectivity using diffusion tractography and mapping cortical diffusion patterns in these problematic regions.

## Discussion

In this work, we developed synergistic approaches to improve gSlider acquisitions where i) reconstruction with $B_1^+$ and $T_1$ corrections effectively mitigate slab-boundary artifacts in short-TR acquisitions, and ii) dynamic MC $B_0$ shimming and high in-plane acceleration achieve an 8-11-fold reduction in $B_0$ distortion. The results demonstrate that the proposed corrections and local-field control can together achieve high-quality, high-fidelity DWI with 1 mm isotropic resolution in 20 minutes on a 3T clinical scanner.

To minimize the slab-boundary artifacts, there is a trade-off between the design of slab-thickness of the RF refocusing pulse and the sensitivity of gSlider to $B_1^+$ inhomogeneity. Using a slightly thicker refocusing slab than the target slab thickness can improve the refocusing performance at the edges of the slab, since the refocusing slab's transition bands are moved away from the target slab region. This is particularly useful in the presence of $B_1^+$ inhomogeneity, where the refocusing performance at the edge of the refocusing slab can degrade significantly (much more than in the excitation pulse). This thicker refocusing approach was employed in our previous work to provide robustness to $B_1^+$ inhomogeneity, where the gSlider acquisition was performed with a TR of ~4.5s. However, the partial excitations in adjacent slabs induced by this broader refocusing pulse can cause striping artifacts for short TR acquisitions due to the partial recovery. For this work, we employed matching excitation and refocusing slab thickness to avoid large $T_1$ recovery issue while correcting for the increased $B_1^+$ inhomogeneity effect through a modified reconstruction. The proposed method corrects both $B_1^+$ inhomogeneity and partial $T_1$ recovery jointly by using a dictionary of RF-encoding matrices, which has been demonstrated to improve the fidelity of gSlider data for short-TR acquisitions.

In this work, a TR of 3.5s was chosen for the acquisition to balance the trade-off between SNR efficiency vs. spin-history and motion sensitivity issues. With our whole-brain 1mm isotropic gSlider acquisition, the TR can be further shortened to 2.5s to provide higher SNR efficiency (44) and shorter total acquisition time. However, at such TR, with a shorter $T_1$ recovery period, slab boundary artifacts and motion sensitivity will be more severe, requiring the development of more advanced correction approach such as in (30). Employing a TR of 3.5s markedly reduced these issues, while still achieving an SNR-efficiently level that is at 80%-95% of the optimal value for white matter and gray matter tissues (44).

There are some limitations in the proposed $B_1^+$ and $T_1$ corrections. First, in the Bloch simulation and dictionary generation, the continuous $B_1^+$ variations were discretized and sorted into several discretized bins, which may not reflect the accurate slab-profile at a given position. However, since the spatial $B_1^+$ values vary slowly and smoothly, the discretization of $B_1^+$ maps should not affect the $B_1^+$ correction unduly. Second, for $T_1$ correction, the $T_1$ value used in the study is 1000ms, which is taken to be the average value across the gray- and white-matter tissues of the brain (45). However, this assumption does not accurately reflect the complicated biochemical environment in the brain, leading to residual slab-boundary artifacts due to partial volume effects caused by high-$T_1$ compartments such as blood and cerebrospinal fluid. Nevertheless, these residual artifacts have minimal impact on diffusion-weighted images because the fluid is almost completely attenuated by the diffusion encodings.

Slab-by-slab dynamic shim updating was used for gSlider acquisition. With a 32-channel custom-built AC/DC coil, the $B_0$ inhomogeneity was reduced by >50%, thereby significantly mitigating the geometric distortion in EPI slices. Combined with parallel imaging with high in-plane acceleration ($R_{inplane}$=4), the geometric distortion was further reduced. Consequently, the resultant images closely matched the TSE reference across all brain regions. Furthermore, we extended our slab-by-slab shimming scheme to slab-group-by-slab-group shimming for SMSb acquisitions that shim two distant slabs simultaneously. Compared to post-processing based EPI distortion correction methods (2,46,47) which require additional data (e.g. reversed phase-encoding images) and processing time, our proposed dynamic MC shimming method only needs a fast low-resolution $B_0$ scan for computing the optimal shim currents and reduce the geometric distortion during the acquisition. Furthermore, for high-resolution DWI, the dynamic MC shimming method reduces geometric distortion *at its source*, whereas the post-processing methods have limited ability to accurately correct for severe voxel pile-up. Dynamic MC shimming is not only applicable to gSlider acquisitions, but can also be easily applied to single- and multi-shot diffusion acquisitions such as single-shot SMS-EPI (33), multi-shot EPI acquisitions (3,48), which can enable high-resolution, high-quality functional MRI and DWI with low geometric distortion and short acquisition times.

## Conclusion

We proposed $B_1^+$ and $T_1$ corrections with the dynamic MC $B_0$ shimming strategy to improve the fidelity of high-resolution gSlider-EPI acquisitions with low geometric distortion. *In vivo* studies demonstrated that the proposed methods enabled markedly improved image quality with reduced slab-boundary artifacts for short-TR acquisition.

The 8-11x reduction in geometric distortion should improve the quality of diffusion data that can be applied to many clinical and neuroscience applications.

## Acknowledgement

This work was supported in part by NIH research grants: R01EB020613, R01MH116173, U01EB025162, P41EB015896, R00EB021349, R01EB016695, and the shared instrumentation grants: S10RR023401, S10RR019307, S10RR019254, and S10RR023043.

**Figure Captions**

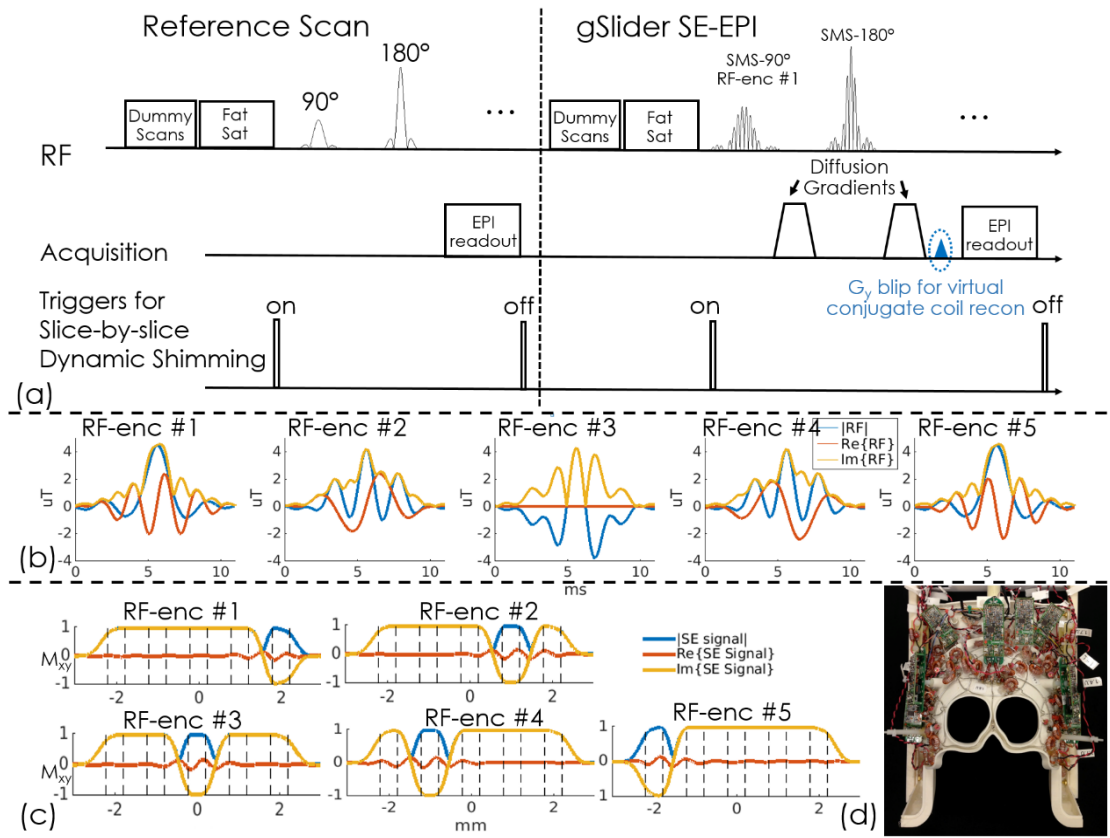

**Figure 1.** (a) Sequence diagram of gSlider acquisition and slab-by-slab triggers. (b) Five RF-encoded gSlider pulses and (c) their corresponding Bloch simulated slab profiles. (d) The 32-channel combined RF receive and $B_0$ shim array ("AC/DC" coil) with independent control of shim currents in each loop in the array.

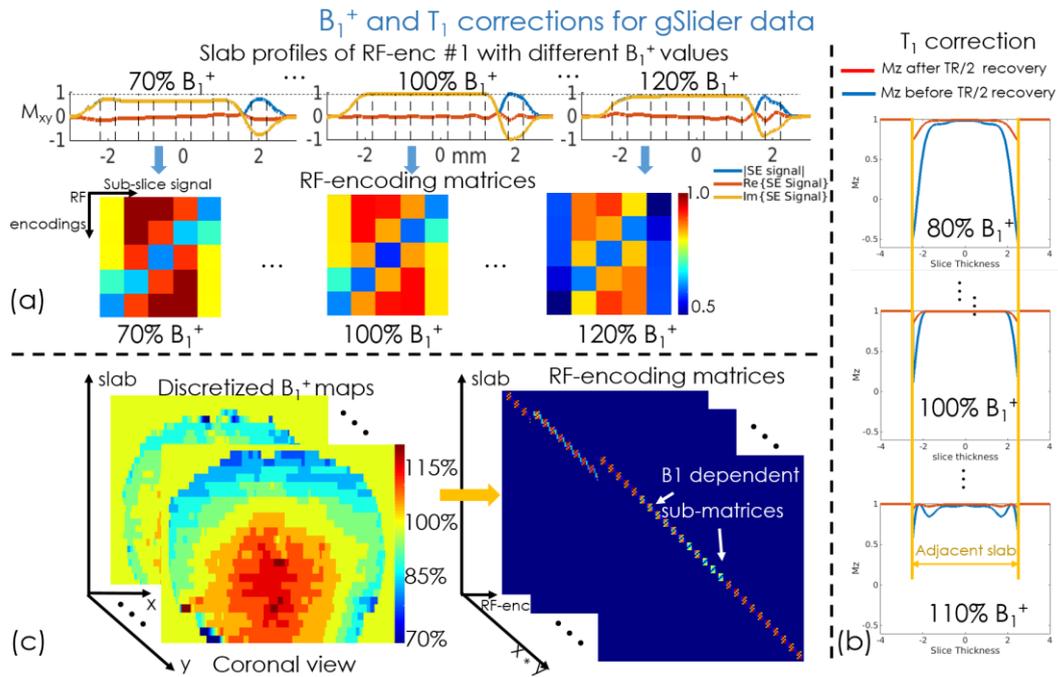

**Figure 2.** (a) Slab profiles ($M_{xy}$) of a RF-encoded gSlider pulse with $B_1^+$ inhomogeneity and (b) Longitude Magnetizations ($M_z$) of the adjacent slab before and after TR/2 recovery. (c) The flowchart of $B_1^+$ correction using the pre-scan $B_1^+$ maps.

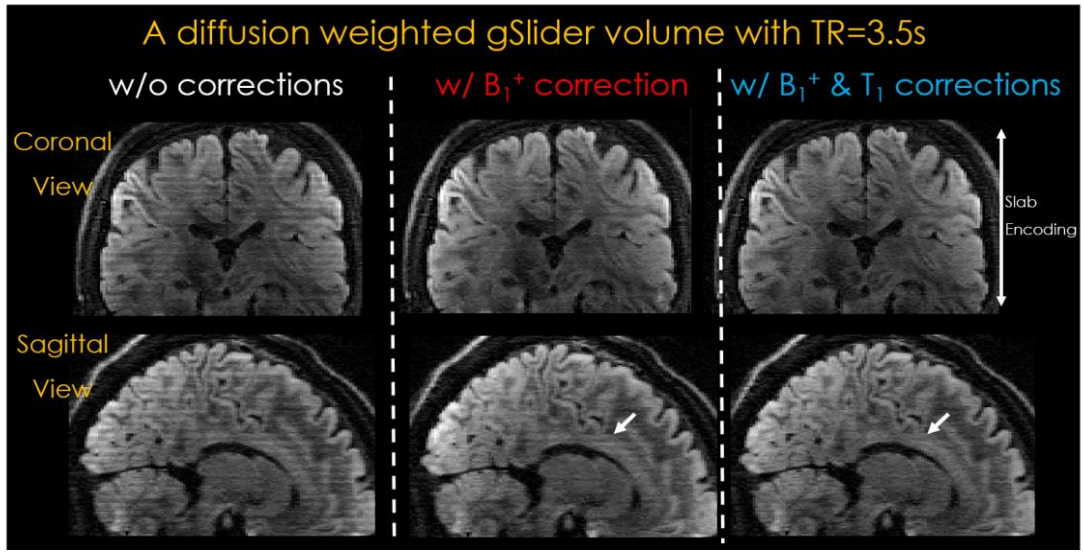

**Figure 3.** gSlider diffusion-weighted image results without correction, with $B_1^+$ correction only, and with $B_1^+$ & $T_1$ corrections. The white arrows highlight the further mitigation of stripping artifacts by using $T_1$ correction. To demonstrate the efficiency of the proposed $B_1^+$ & $T_1$ correction methods, the diffusion weighted gSlider data were acquired in axial orientation, with resolution of $1.5 \times 1.5 \times 1.0$ mm$^3$.

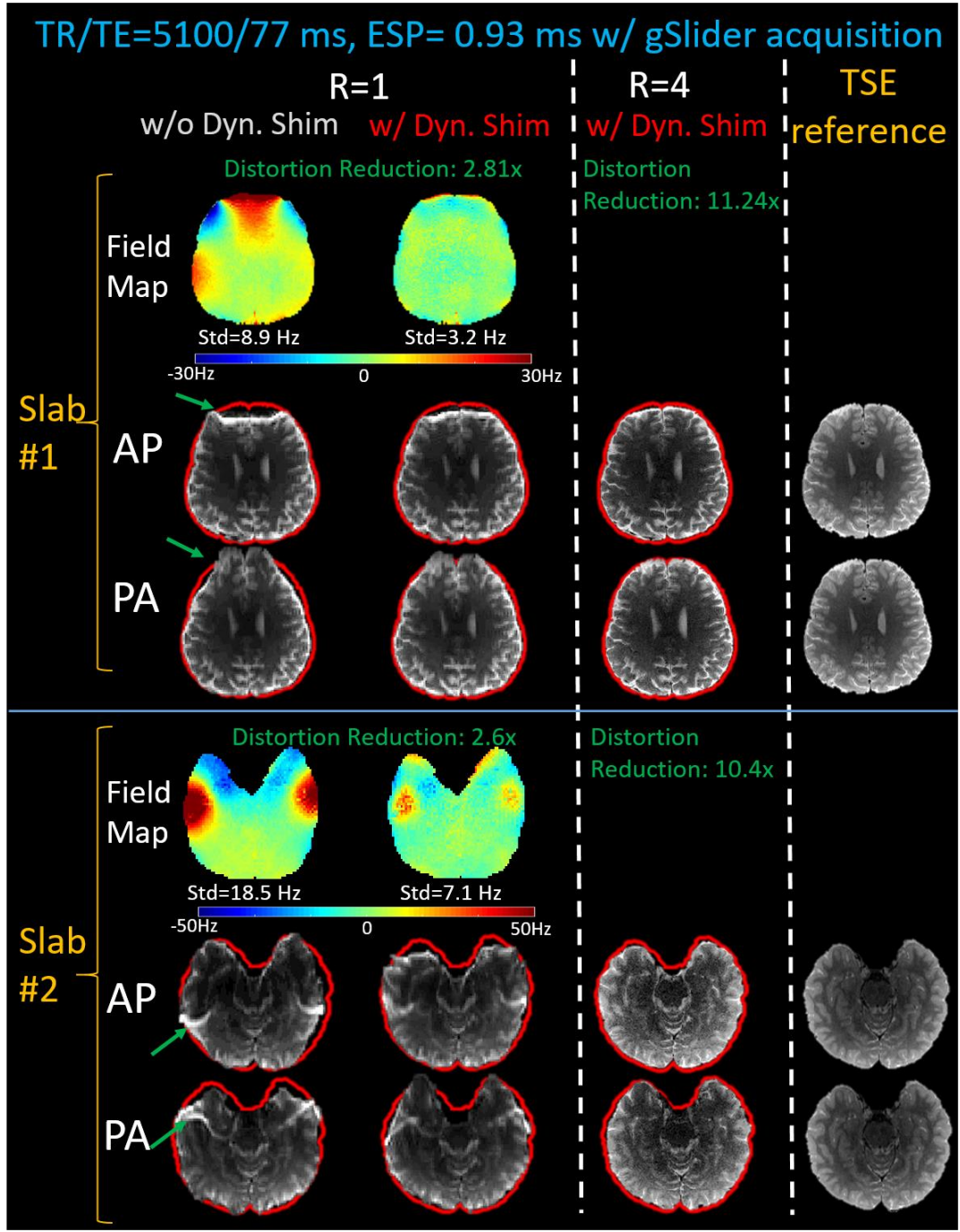

**Figure 4.** The comparisons of the $B_0$ field maps and MB-1 EPI distortions with Anterior-to-Posterior (A-P) and Posterior-to-Anterior (P-A) phase-encodings, with and without dynamic multi-coil $B_0$ shimming. The shimming is then combined with $R_{inplane}=4$ acceleration to achieve ~10-fold reduction in EPI distortion, bringing the A-P and P-A images into closer alignment. The resulting low-distortion EPI slices resemble the distortion-free $T_2$-TSE reference slices (see red brain mask outline).

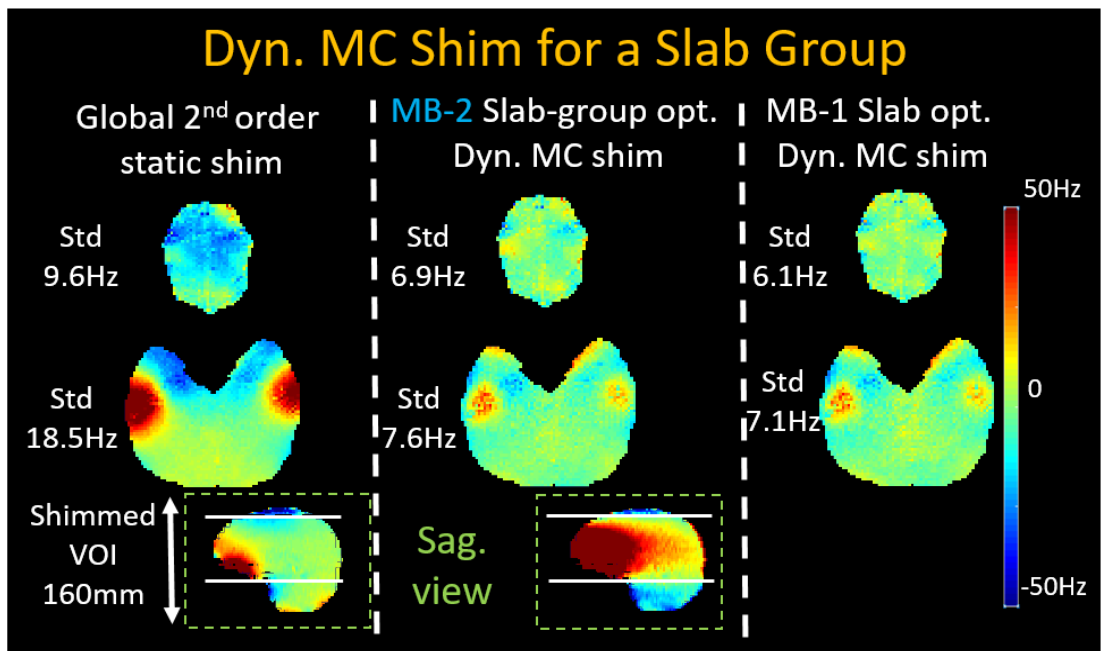

**Figure 5.** Compared to baseline 2$^{nd}$-order spherical harmonic global shimming, dynamic MC shimming for MB-2 acquisitions jointly shims two slabs at the same time, while out-of-slab regions are unconstrained and are allowed to have poor shims. The shim results of MB-2 are performance closely to MB-1 slab-optimized shimming.

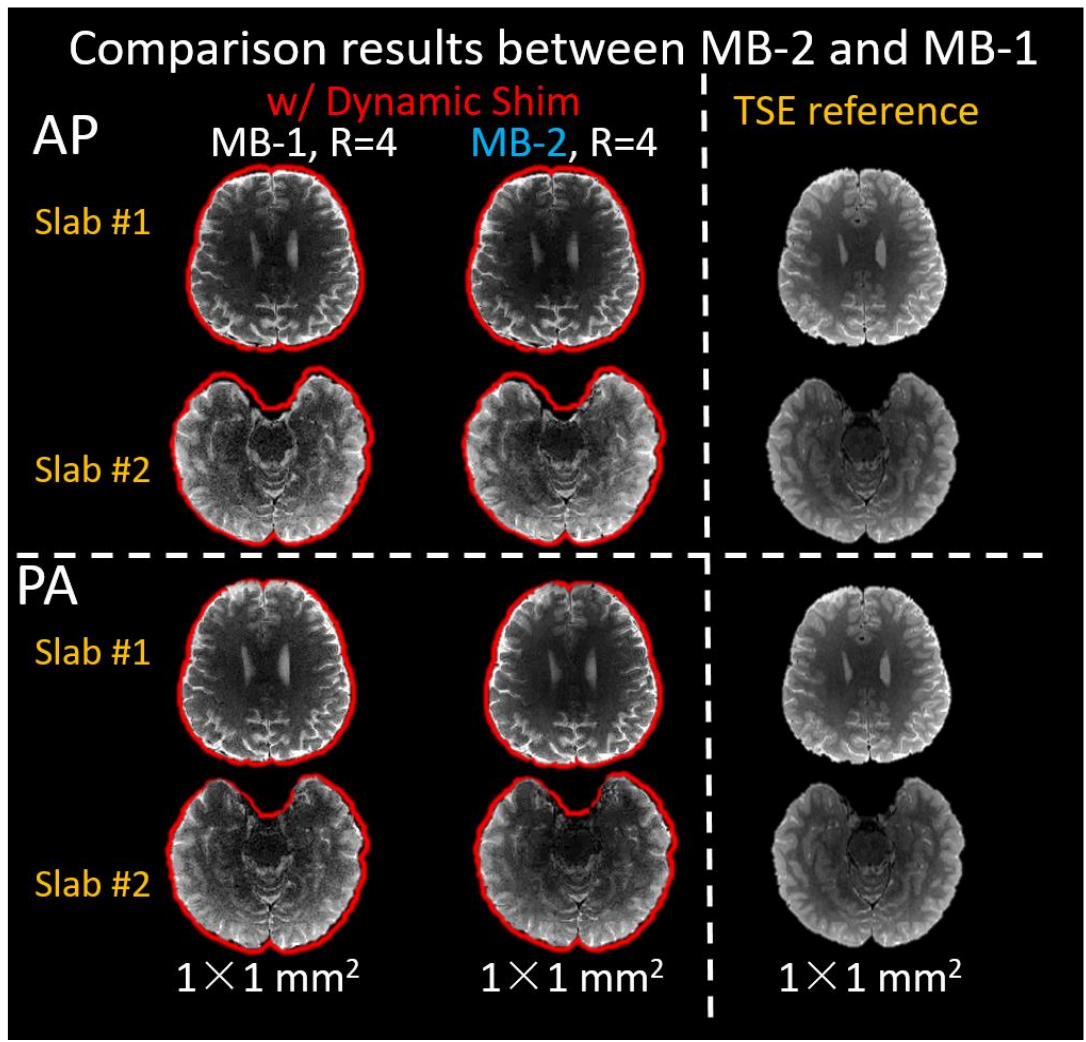

**Figure 6.** Comparison between MB-2 and MB-1. The MB-2 shimming performance closely resembles that of the MB-1 slab-optimized shimming. The red outlines show the contours of $T_2$-TSE reference brain mask.

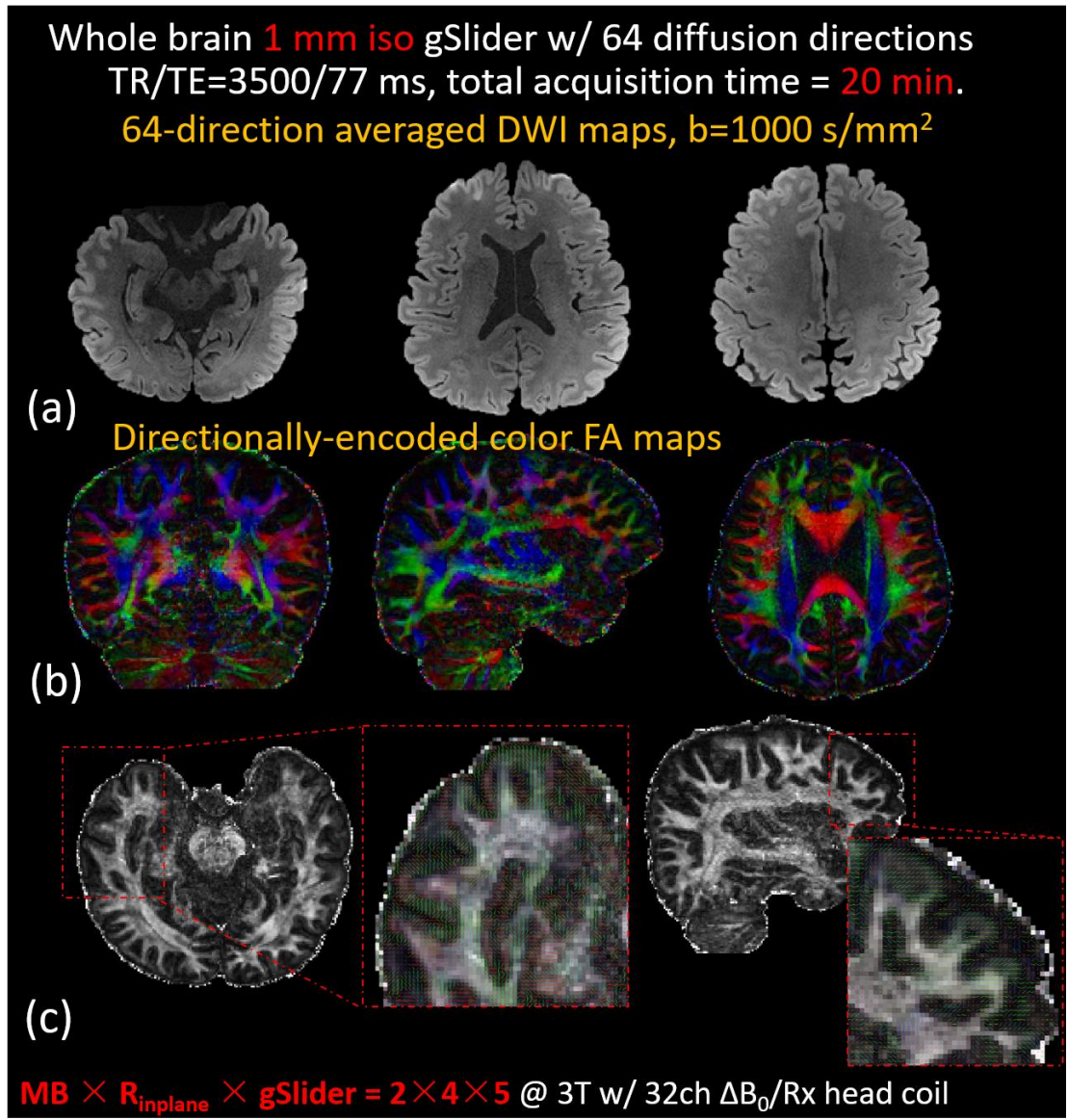

**Figure 7.** (a) Averaged diffusion-weighted images and (b) directionally-encoded color fractional anisotropy (FA) maps of the 1 mm isotropic diffusion-weighted data with 64 directions obtained in ~20-minutes using the proposed correction and dynamic shimming framework. (c) The primary eigenvectors from diffusion tensor imaging in typically problematic temporal and frontal lobes of the brain were more accurately depicted by synergistically combining dynamic $B_0$ shimming with parallel imaging acceleration. The primary eigenvectors were color-encoded (red: left-right, green: anterior-posterior, blue: superior-inferior) and overlaid on FA maps.